\documentclass[11pt, letterpaper]{article}

\usepackage{ifthen}

\newcommand{\MyMacro}[4]{
\newboolean{#1}
\setboolean{#1}{#2}
\newcommand{#3}{\ifthenelse{\boolean{#1}}}
\newcommand{#4}{\ifthenelse{\not \boolean{#1}}}
}

\usepackage{ifthen}

\ifthenelse{\isundefined{\MyMacro}}{

\newcommand{\MyMacro}[4]{
\newboolean{#1}
\setboolean{#1}{#2}
\newcommand{#3}{\ifthenelse{\boolean{#1}}}
\newcommand{#4}{\ifthenelse{\not \boolean{#1}}}
}

}

\newcommand{\ifndef}[2]{\ifthenelse{\isundefined{#1}}{#2}{}}

\newcommand{\mydef}[2]{\def#1{#2}}

\newcommand{\nospell}[1]{#1}   %

\usepackage{amssymb}
\usepackage{amsmath}   %

\usepackage{amsthm}    %
\usepackage{latexsym}
\usepackage{amsfonts}
\usepackage{units}     %
\usepackage{psfrag}    %
\usepackage{url}    %

\usepackage[dvips]{graphicx}
\usepackage[usenames,dvipsnames]{color}

\newcommand{\MyComment}[1]{\ClassWarning{My Macros}{#1}}

\ifndef{\theorem}{\newtheorem{theorem}{Theorem}[section]}
\ifndef{\lemma}{\newtheorem{lemma}[theorem]{Lemma}}
\ifndef{\corollary}{\newtheorem{corollary}[theorem]{Corollary}}
\ifndef{\remark}{\theoremstyle{remark} \newtheorem{remark}{Remark}}
\ifndef{\proposition}{\newtheorem{proposition}{Proposition}}
\newtheoremstyle{mydefinition}   %
{\topsep}{\topsep}   %
{\slshape}   %
{}   %
{\bfseries}   %
{.}   %
{ }   %
{}   %
{\theoremstyle{mydefinition}}
\newtheoremstyle{myexample}   %
{\topsep}{\topsep}   %
{\itshape}   %
{}   %
{\slshape}   %
{:}   %
{ }   %
{\ul{\thmname{#1}}}   %
\ifndef{\example}{\theoremstyle{myexample} \newtheorem{example}{Example}}
\ifndef{\claim}{\newtheorem{claim}[theorem]{Claim}}
\ifndef{\result}{\newtheorem{result}[theorem]{Result}}
\ifndef{\problem}{\newtheorem{problem}{Problem}}
\ifndef{\protocol}{\newtheorem{protocol}{Protocol}}
\newtheoremstyle{myclaims}   %
{\topsep}{\topsep}   %
{\slshape}   %
{}   %
{\bfseries\itshape}   %
{}   %
{ }   %
{\thmname{#1}\thmnumber{ \!#2}.}   %
{\theoremstyle{myclaims}

\ifndef{\fact}{\newtheorem{fact}[theorem]{Fact}}
\ifndef{\observation}{\newtheorem{observation}[theorem]{Observation}}
}

\newtheoremstyle{anystatement}{\topsep}{\topsep}{\itshape}{}{\bfseries}{.}{ }{\anystatementname}
{\theoremstyle{anystatement}}

\newcommand{\anystatementname}{}

\newcounter{tmp_id_cnt}
\newcommand{\AuxNew}[4][]{#2{#3}[1][*]%
{\ifthenelse{\equal{*}{##1}} %
{\Ensuremath{#1{#4}}}%
{\ifthenelse{\equal{b}{##1}} %
{\Ensuremath{\mathbf{#4}}}%
{\ifthenelse{\equal{}{##1}} %
{\IfMathMode{#1{#4}}{#4}}{}}}}}

\newcommand{\newident}[3][*]{\ifthenelse{\equal{*}{#1}}%
{\AuxNew[\mathit]{\newcommand}{#2}{#3}} %
{\mydef{#2}{\Ensuremath{\mathit{#3}}}}} %

\newcommand{\newmat}[3][*]{\ifthenelse{\equal{*}{#1}}%
{\AuxNew{\newcommand}{#2}{#3}} %
{\mydef{#2}{\Ensuremath{#3}}}} %

\newcommand{\providemat}[3][*]{\ifthenelse{\equal{*}{#1}} %
{\AuxNew{\providecommand}{#2}{#3}} %
{\mydef{#2}{\Ensuremath{#3}}}} %

\newcommand{\MyMakeTheoMacros}[3]{
\newcommand{#2}[2][]{\ifthenelse{\equal{}{##1}}
{\begin{#1} ##2 \end{#1}}
{\begin{#1}\label{##1} ##2\end{#1}}}
\newcommand{#3}[3][]{\ifthenelse{\equal{}{##1}}
{\begin{#1}{\e{##2}} ##3 \end{#1}}
{\begin{#1}{\e{##2}}\label{##1} ##3\end{#1}}}
}

\newcommand{\MyMakeDupTheoMacros}[8]{
\MyMakeTheoMacros{#1}{#2}{#3}
\newcommand{#4}[3]{
\newcommand{##2}{##3}
\begin{#1}\label{##1} ##2\end{#1}}
\newcommand{#5}[4]{
\newcommand{##2}{##4}
\begin{#1}{\e{##3}}\label{##1} ##2\end{#1}}
\newcommand{#8}[2]{\def\my_tmp_id{my_tmp_id_\arabic{tmp_id_cnt}}
\newtheorem*{\my_tmp_id}{#7~\ref{##1}}
\begin{\my_tmp_id} ##2 \end{\my_tmp_id}\stepcounter{tmp_id_cnt}}
\newcommand{#6}[6]{
#2[##1]{##2}

##3
\prf[#7~\ref{##1}]{##6} \newcommand{##5}{}

}
}

\newcommand{\MyMakeRefMacros}[3]{\newcommand{#1}[2][]
{\ifthenelse{\equal{}{##1}}{#2~\ref{##2}}{#3~\ref{##1} and~\ref{##2}}}}

\newcommand{\MyMakeEqRefMacros}[3]{\newcommand{#1}[2][]
{\ifthenelse{\equal{}{##1}}{#2~\eqref{##2}}{#3~\eqref{##1} and~\eqref{##2}}}}

\newcommand{\abstr}[1]{
\begin{abstract}
#1
\end{abstract}}

\newcommand{\bibentry}[8]{

\bibitem[\nospell{#8}]{#1} {\textup #3}.
\ifthenelse{\equal{}{#6}}
{\newblock \textrm{#4.} \newblock {\em #5}, #7.}
{\newblock \textrm{#4.} \newblock {\em #5, #6}, #7.}
}

\MyMakeTheoMacros{fact}{\fct}{\nfct}

\MyMakeRefMacros{\fctref}{Fact}{Facts}

\MyMakeTheoMacros{observation}{\obs}{\nobs}

\MyMakeRefMacros{\obsref}{Observation}{Observations}

\MyMakeDupTheoMacros{lemma}
{\lem}{\nlem}{\lemdup}{\nlemdup}{\lemapp}{Lemma}{\lemrep}

\MyMakeRefMacros{\lemref}{Lemma}{Lemmas}

\MyMakeDupTheoMacros{corollary}
{\crl}{\ncrl}{\crldup}{\ncrldup}{\crlapp}{Corollary}{\crlrep}

\MyMakeRefMacros{\crlref}{Corollary}{Corollaries}

\MyMakeTheoMacros{proposition}{\prp}{\nprp}

\newtheorem*{prp*}{\e{Proposition}}

\MyMakeRefMacros{\prpref}{Proposition}{Propositions}

\MyMakeDupTheoMacros{my_claim}
{\clm}{\nclm}{\clmdup}{\nclmdup}{\clmapp}{Claim}{\clmrep}

\MyMakeRefMacros{\clmref}{Claim}{Claims}

\MyMakeDupTheoMacros{theorem}
{\theo}{\ntheo}{\theodup}{\ntheodup}{\theoapp}{Theorem}{\theorep}

\MyMakeRefMacros{\theoref}{Theorem}{Theorems}

\MyMakeTheoMacros{definition}{\defi}{\ndefi}

\MyMakeRefMacros{\defiref}{Definition}{Definitions}

\MyMakeTheoMacros{problem}{\prob}{\nprob}

\MyMakeRefMacros{\probref}{Problem}{Problems}

\MyMakeTheoMacros{protocol}{\prot}{\nprot}

\MyMakeRefMacros{\protref}{Protocol}{Protocols}

\providecommand{\qedsymbol}{\square}

\newcommand{\prf}[2][]{\ifthenelse{\equal{}{#1}}%
{\begin{proof}\renewcommand{\qedsymbol}{$\blacksquare$}%
#2 \end{proof}}%
{\begin{proof}[Proof of #1]%
\renewcommand{\qedsymbol}{$\blacksquare_{\mbox{\it{\scriptsize{#1}}}}$}%
#2 \end{proof}}
}

\newcommand{\sect}[2][]{\ifthenelse{\equal{}{#1}}
{\section{#2}}
{\section{#2}\label{#1}}}

\newcommand{\ssect}[2][]{\ifthenelse{\equal{}{#1}}
{\subsection{#2}}
{\subsection{#2}\label{#1}}}

\MyMakeRefMacros{\chref}{Chapter}{Chapters}

\MyMakeRefMacros{\sref}{Section}{Sections}

\MyMakeRefMacros{\ssref}{Subsection}{Subsections}

\MyMakeRefMacros{\sssref}{Subsection}{Subsections}

\definecolor{DarkGreen}{rgb}{0,0.45,0.08}
\definecolor{LightBlue}{rgb}{0.122,0.016,0.855}

\newcommand{\IfMathMode}[2]{\ifmmode{#1}\else{#2}\fi}

\newcommand{\Ensuremath}{\ensuremath}

\newcommand{\fbr}[1]{\IfMathMode %
{#1}{$#1$}}                      %

\newcommand{\fnbr}[1]{\mbox{\fbr{#1}}}   %

\newcommand{\fla}[2][*]{\ifthenelse{\equal{}{#1}}{\fbr{#2}}{\fnbr{#2}}}

\newcommand{\mat}[2][]{\ifthenelse{\equal{}{#1}} %
{ \begin{displaymath} #2 \end{displaymath} } %
{ \begin{equation} \label{#1} #2 \end{equation} }%
}

\newcommand{\matal}[2][]{\mat[#1]{\begin{aligned} #2 \end{aligned}}}

\newcommand{\m}{\mat}
\newcommand{\mal}{\matal}

\MyMakeEqRefMacros{\equref}{Equation}{Equations}

\MyMakeEqRefMacros{\expref}{Expression}{Expressions}

\MyMakeEqRefMacros{\inequref}{Inequality}{Inequalities}

\MyMakeRefMacros{\figref}{Figure}{Figures}

\providecommand{\middle}{\big}

\newcommand{\h}[2][]{\ifthenelse{\equal{}{#2}}%
{\mathop{\mathbf{H}}_{#1}}%
{\mathop{\mathbf{H}}_{#1}{\left[{#2}\right]}}}
\newcommand{\hh}[3][]{\mathop{\mathbf{H}}_{#1}%
{\left[{#2}\middle|\vphantom{|_1^1}{#3}\right]}}

\newcommand{\KL}[2]{D\llp{#1}\middle|\middle|{#2}\rrp}

\providecommand{\E}[2][]{\mathop{\mathbf{E}}_{#1}{\left[{#2}\right]}}

\newcommand{\PR}[2][]{\mathop{\mathbf{Pr}}_{#1}{\left[{#2}\right]}}

\newcommand{\pl}[1][]{\nospell{\ifthenelse{\equal{}{#1}}%
{\!\stackrel'{}\!\!\txt{s}}%
{\fla{#1\!\stackrel'{}\!\!\txt{s}}}}}

\newmat[]{\llp}{\left(}
\newmat[]{\rrp}{\right)}

\newmat[]{\tm}{\cdot}
\newmat[]{\sbseq}{\subseteq}
\newmat[]{\deq}{\stackrel{\textrm{def}}{=}}

\providemat{\QQ}{\mathbb{Q}}

\newmat[]{\ds}{\dots}
\newmat[]{\dc}{,\ds,}
\newmat[]{\dtm}{\times\ds\times}

\newcommand{\fr}[3][*]{%
\ifthenelse{\equal{*}{#1}}        %
{\frac{#2}{#3}}{}%
\ifthenelse{\equal{/}{#1}}        %
{\nicefrac{#2}{#3}}{}%
\ifthenelse{\equal{}{#1}}         %
{\left.#2\middle/#3\right.}{}%
\ifthenelse{\equal{p_}{#1}}       %
{\left.\left(#2\right)\middle/#3\right.}{}%
\ifthenelse{\equal{_p}{#1}}       %
{\left.#2\middle/\left(#3\right)\right.}{}%
\ifthenelse{\equal{pp}{#1}}       %
{\left.\left(#2\right)\middle/\left(#3\right)\right.}{}
}

\def\MySQRT#1#2{    %
\setbox0=\hbox{$#1\sqrt{#2\,}$}\dimen0=\ht0%
\advance\dimen0-0.2\ht0%
\setbox2=\hbox{\vrule height\ht0 depth -\dimen0}%
{\box0\lower0.4pt\box2}}

\newcommand{\set}[2][]{\ifthenelse{\equal{}{#1}} %
{\Ensuremath{\left\{#2\right\}}}%
{\Ensuremath{\left\{#2\,\middle\arrowvert\,#1\right\}}}}

\newcommand{\newfunction}[2]{ %
\newcommand{#1}[2][*]{\ifthenelse{\equal{*}{##1}}%
{\Ensuremath{#2{\left(##2\right)}}}%
{#2(##2)}}%
}

\renewcommand{\l}{\left}
\renewcommand{\r}{\right}
\newcommand{\eps}{\varepsilon}

\newcommand{\sz}[2][]{\ifthenelse{\equal{}{#1}}%
{\Ensuremath{\left|#2\right|}}%
{\Ensuremath{\left|#2\middle|_{#1}\right.}}}

\newcommand{\e}{\emph}

\newcommand{\ul}[1]{\underline{#1}}  %

\newcommand{\txt}[1]{\textrm{#1}}   %

\DeclareMathAlphabet{\lowcal}{OT1}{pzc}{m}{it}

\usepackage{hyphenat}   %

\date{}
\title{A Tail Bound for Read-$k$ Families of Functions}

\author{
{\bf Dmitry Gavinsky}\\
{\small NEC Laboratories America, Inc.}\\
{\small Princeton, NJ 08540, U.S.A.}
\and
{\bf Shachar Lovett}\\
{\small Institute of Advanced Study}\\
{\small Princeton, NJ 08540, U.S.A.}
\and
{\bf Michael Saks}\\
{\small Department of Mathematics, Rutgers University}\\
{\small Piscataway, NJ 08854, U.S.A.}
\and
{\bf Srikanth Srinivasan}\\
{\small DIMACS, Rutgers University}\\
{\small Piscataway, NJ 08854, U.S.A.}
}

\begin{document}

\maketitle

\thispagestyle{empty}

\abstr{We prove a Chernoff-like large deviation bound on the sum of non-independent random variables that have the following dependence structure.
The variables $Y_1,\ldots,Y_r$ are arbitrary Boolean functions of independent random variables $X_1,\ldots,X_m$, modulo a restriction that every $X_i$ influences at most $k$ of the variables $Y_1,\ldots,Y_r$.}

\setcounter{page}{1}

\sect{Introduction}

Let $Y_1,\ldots,Y_r$ be independent indicator random variables with $\PR{Y_i=1}=p$ and let $Y=Y_1+\ldots+Y_r$ denote their sum. The average number
of variables which are set is $\E{Y}=pr$. A basic question is how concentrated is the sum around its average. The answer is given by Chernoff bound: for any $\eps>0$, the probability that the fraction of variables that occur exceeds the expectation by more than $\eps$ is bounded by
$$
\PR{Y_1+\ldots+Y_r \ge (p+\eps) r} \le e^{-\KL{p+\eps}{p} \cdot r} \le e^{-2 \eps^2 r},
$$
and similarly, the probability that it is below the expectation by more than $\eps$ is bounded by
$$
\PR{Y_1+\ldots+Y_r \le (p-\eps) r} \le e^{-\KL{p-\eps}{p} \cdot r} \le e^{-2 \eps^2 r}.
$$
Here, $\KL{q}{p}$ is Kullback-Leibler divergence defined as
$$
\KL{q}{p}=q \log\left(\frac{q}{p}\right)+(1-q) \log\left(\frac{1-q}{1-p}\right),
$$
and the latter estimate (which is the one more commonly used in applications) follows from $\KL{q}{p} \ge 2 (q-p)^2$.

Our focus in this paper is on deriving similar estimates when the variables $Y_1,\ldots,Y_r$ are not independent, but the level of dependence between them is in some sense 'weak'. These type of questions are commonly studied in probability theory, as they allow one to apply Chernoff-like bounds to more general scenarios. For example, one case which is used in various applications is where the variables $Y_1,\ldots,Y_r$ are assumed to be $k$-wise independent. In this case, Bellare and Rompel~\cite{BellareRompel94} obtained Chernoff-like tail bounds on the probability that the number of set variables deviates from its expectation. Another well studied case is when $Y_1,\ldots,Y_r$ form a martingale, in which case Azuma inequality and its generalizations give bounds which are comparable to Chernoff bound.

We consider in this paper another model of weak dependence. Assume that the variables $Y_1,\ldots,Y_r$ can be factored as functions of independent random variables $X_1,\ldots,X_m$. More concretely, each $Y_j$ is a function of a subset of the variables $X_1,\ldots,X_m$.
One extreme case is that these subsets are disjoint, in which case the variables $Y_1,\ldots,Y_r$ are independent. The other extreme case is when these subsets all share a common element, in which case $Y_1,\ldots,Y_r$ can be arbitrary. We say that $Y_1,\ldots,Y_r$ are a {\em read-$k$ family} if there exists such a factorization where each $X_i$ influences at most $k$ of the variables $Y_1,\ldots,Y_r$.

\ndefi{(Read-$k$ families)}{
Let $X_1\dc X_m$ be independent random variables. For $j \in [r]$, let $P_j\sbseq[m]$ and let $f_j$ be a Boolean function of $\l(X_i\r)_{i\in P_j}$. Assume that $\sz{\set[i\in P_j]j}\le k$ for every $i\in [m]$. Then the random variables $Y_j=f_j((X_i)_{i \in P_j})$ are called a read-$k$ family.
}

There are several motivations for studying read-$k$ families of functions: for example, they arise naturally when studying subgraphs counts in random graphs; or in generalizations of read-once models in computational complexity. We will not discuss applications in this paper, but instead focus on the basic properties of read-$k$ families. Our main result is that Chernoff-like tail bounds hold for read-$k$ families, where the bounds degrade as $k$ increases.

\theo[t_intro_tail]{
Let $Y_1,\ldots,Y_r$ be a family of read-$k$ indicator variables with $\PR{Y_i=1}=p_i$, and let $p$ be the average of $p_1,\ldots,p_r$. Then for any $\eps>0$,
$$
\PR{Y_1+\ldots+Y_r \ge (p+\eps) r} \le e^{-\KL{p+\eps}{p} \cdot r/k}
$$
and
$$
\PR{Y_1+\ldots+Y_r \le (p-\eps) r} \le e^{-\KL{p-\eps}{p} \cdot r/k}.
$$
}
That is, we obtain the same bound as that of the standard Chernoff bound, except that the exponent is divided by $k$. This is clearly tight: let $Y_1=\ldots=Y_k=X_1$, $Y_{k+1}=\ldots=Y_{2k}=X_2$, etc, where $X_1,X_2,\ldots,X_{r/k}$ are independent with $\Pr[X_i=1]=p$. Then for example
$$
\PR{Y_1+\ldots+Y_r \ge (p+\eps) r} = \PR{X_1+\ldots+X_{r/k} \ge (p+\eps) r/k}.
$$

We note that concentration bounds for $Y = \sum_i Y_i$ may also be obtained by observing that $Y$ is a $k$-Lipschitz function of the underlying independent random variables $X_1,\ldots,X_m$ and then applying standard Martingale-based arguments, such as Azuma's inequality (see, for example, \cite{DubhashiPanconesi}). However, these techniques are limited in the sense that they only yield concentration bounds when the deviation $\eps\cdot r$ above is at least $\sqrt{m}$ or so, whereas our results yield concentration bounds even when the deviation is roughly the square root of the mean $p\cdot r$, which may be much smaller. It is also known (see \cite[Section 4]{Vondraksurvey}) that such `dimension-free' concentration bounds cannot be obtained for Lipschitz functions in general.

\ssect{Proof overview}

Let $Y_1,\ldots,Y_r$ be a read-$k$ family with $\Pr[Y_i=1]=p_i$. Let us consider for a moment a more basic question: what is the maximal probability that $Y_1=\ldots=Y_r=1$? The answer to this question is given by Shearer's lemma \cite{ChungGrFrSh86}, which is a simple yet extremely powerful tool.

\theo[t_intro_shearer]{
Let $Y_1,\ldots,Y_r$ be a family of read-$k$ indicator variables with $\PR{Y_i=1}=p$. Then
$$
\PR{Y_1=\ldots=Y_r=1} \le p^{r/k}.
$$
}
Note that the answer is the $k$-th root of the answer in the case where the variables are independent, similar to what we get for the tail bounds. The result is tight, as can be seen by taking $Y_1=\ldots=Y_k=X_1$, $Y_{k+1}=\ldots=Y_{2k}=X_2$, etc, where $X_1,X_2,\ldots,X_{r/k}$ are independent with $\PR{X_i=1}=p$. The proof of Shearer's lemma is based on the entropy method.
We refer the interested reader to a survey of Radhakrishnan~\cite{Radhakrishnan2003} on applications of Shearer's lemma; and to continuous analogs by
Finner~\cite{Finner92} and Friedgut~\cite{Friedgut2004}.

We derive \theoref{t_intro_tail} by constructing an information-theoretic proof of Chernoff bound and applying Shearer's lemma to make the proof robust for the case of non-independent random variables forming a read-$k$ family.
In our proof we use the ``entropy method'' that was introduced by Ledoux~\cite{L96_On_Tal}, and more recently has been used to prove a number of concentration inequalities (cf.~\cite{BLM03_Con_In}).
From a technical point of view, we construct analogs of Shearer's lemma for Kullback-Leibler divergence.

\sect{Preliminaries}

Let $X$ be a random variable and $A$ be a subset of its support. We write ``$X\in A$'' to address the event ``the value of $X$ belongs to $A$''.

\ssect{Entropy}

If $X$ is distributed according to $\mu$, then we will write both $\h X$ and $\h\mu$ to denote the entropy $\E[X=x]{\log(1/\mu(x))}$. All logarithms in this paper are natural. If $X$ is supported in the set $A$ then $\h X \le \log |A|$,
where equality holds when $X$ is the uniform distribution over $A$. For two random variables $X,Y$ we denote their conditional entropy by $\h{X|Y}=\E[Y=y]{\h{X|Y=y}}$. We note that always $\h{X} \ge \h{X|Y}$. If $X_1,\ldots,X_m$ are random variables then $\h{X_1,\ldots,X_m}$ denotes the entropy of their joint distribution. The chain rule of entropy is
$$
\h{X_1,\ldots,X_m}=\h{X_1}+\h{X_2|X_1}+\ldots+\h{X_m|X_1,\ldots,X_{m-1}}.
$$

\ssect{Relative entropy}

Let $\mu'$ and $\mu$ be distributions defined over the same discrete domain $A$. The relative entropy (or Kullback-Leibler divergence) between $\mu'$ and $\mu$ is defined as
\m{\KL{\mu'}{\mu}\deq
\sum_{a\in A}\mu'(a)\log\fr{\mu'(a)}{\mu(a)}.}
If random variables $X,X'$ are distributed according to $\mu,\mu'$, accordingly, then we write $\KL{X'}{X} = \KL{\mu'}{\mu}$.
Kullback-Leibler divergence also satisfies a chain rule. We will only need the following simple corollary of it.

\fct[f_KL_monotone]{Let $X,X'$ be random variables defined on a domain $A$. Let $\phi:A \to B$ be some function. Then
\m{\KL{X'}{X} \ge \KL{\phi(X')}{\phi(X)}.}
}

We will usually use relative entropy in the case when the second operand is the uniform distribution over some set.
The following observations will be useful.

\fct[f_KLu]{Let $\mu$ be the uniform distribution over some set $A$, and let $\mu'$ be any distribution over $A$. Then
\m{\KL{\mu'}\mu=\h\mu-\h{\mu'}.}
}

\clm[c_E2KL]{Let $\mu$ be the uniform distribution over some set $A$, and let $A'\sbseq A$ satisfy $\mu(A')=p$.
Let $\mu'$ be any distribution over $A$, satisfying $\mu'(A')=q$.
Then
\m{\KL{\mu'}\mu\ge\KL qp.}}

Here we use the standard convention $\KL qp$ to denote the relative entropy between Bernoulli distributions whose probabilities of positive outcomes are, respectively, $q$ and $p$.

\prf{
Let $X$ be a random variable taking values in $A$ according to the distribution $\mu'$, and let $I$ be the indicator of the event ``$X\in A'$''.
Let $\h q$ denote the entropy of Bernoulli distribution whose probability of positive outcome is $q$, then
\mal{\h{\mu'}
&=\h I+\hh XI\le
\h q+q\log\l(\sz{A'}\r)+(1-q)\log\l(\sz A-\sz{A'}\r)\\
&=\h q+q\log\l(p\sz A\r)+(1-q)\log\l((1-p)\sz A\r)\\
&=\h q+\log\sz A+q\log p+(1-q)\log(1-p)\\
&=\h{\mu}-\KL qp,}
and the result follows by \fctref{f_KLu}.}

We will need the following facts on $\KL qp$.
The first
is the convexity of Kullback-Leibler divergence.
The last two relate to monotonicity of Kullback-Leibler divergence.

\fct[f_KL_convex]{Let $0 \le p_1,p_2,q_1,q_2,\lambda \le 1$. Then
\m{\KL{\lambda p_1 + (1-\lambda) p_2}{\lambda q_1 + (1-\lambda) q_2} \le \lambda \cdot \KL{p_1}{q_1}+(1-\lambda) \cdot \KL{p_2}{q_2}.}
}

\fct[f_KL_increasing]{Let $0 \le p \le q \le q' \le 1$. Then $\KL{q'}{p} \ge \KL{q}{p}$.}
\fct[f_KL_decreasing]{Let $0 \le q' \le q \le p \le 1$. Then $\KL{q'}{p} \ge \KL{q}{p}$.}

\ssect{Shearer's Lemma}

In order to derive bounds on mutual entropy we will use an elegant tool of amazing universality, \e{Shearer's Lemma}~\cite{ChungGrFrSh86}.
The following formulation of the lemma will be convenient for us.

\nlem[l_Shear]{[Shearer's Lemma]}{Let $X_1\dc X_m$ be random variables and let $P_1\dc P_r\sbseq[m]$ be subsets such that for every $i\in [m]$ it holds that $\sz{\set[i\in P_j]j}\ge k$.
Then
\m{k\tm\h{(X_1\dc X_m)}\le\sum_{j=1}^r\h{(X_i)_{i\in P_j}}.}
}

For completeness we include a short proof, based on the idea commonly attributed to Jaikumar Radhakrishnan.

\prf{Let us denote $s_j = |P_j|$ and let $P_j=\{i_{j,1},\ldots,i_{j,s_j}\}$, where we order the elements $i_{j,1} < \ldots < i_{j,s_j}$. We apply the chain rule for entropy:
\m{\h{(X_i)_{i\in P_j}}=
\sum_{t=1}^{s_j}\hh{X_{i_{j,t}}}{X_{i_{j,1}}\dc X_{i_{j,t-1}}}\ge
\sum_{t=1}^{s_j}\hh{X_{i_{j,t}}}{X_1\dc X_{i_{j,t}-1}},}
where the second inequality follows from non-increasing of entropy as a result of conditioning.
Summing over all $1 \le j \le r$ we obtain that
\m{\sum_{j=1}^r\h{(X_i)_{i\in P_j}}\ge
\sum_{i=1}^m\sz{\set[i\in P_j]j}\tm\hh{X_i}{X_1\dc X_{i-1}}\ge
k\tm\sum_{i=1}^m\hh{X_i}{X_1\dc X_{i-1}},}
and the result follows.}

We need the following simple corollary of Shearer's Lemma:

\crl[c_Shear]{Let $X_1\dc X_m$ be random variables and let $P_1\dc P_r\sbseq[m]$ be such that for every $i\in [m]$ it holds that $\sz{\set[i\in P_j]j}\le k$.
For each $i\in[m]$, let $Y_i$ be an independent random variable, uniformly distributed over a set that includes the support of $X_i$.
Then
\m{k\tm\KL{(X_i)_{i=1}^m}{(Y_i)_{i=1}^m}\ge
\sum_{j=1}^r\KL{(X_i)_{i\in P_j}}{(Y_i)_{i\in P_j}}.}
}

Note the difference in the conditions put upon $\sz{\set[i\in P_j]j}$:\ in Shearer's Lemma it was ``$\ge k$'', and in the corollary it is ``$\le k$''.

\prf{First observe that w.l.o.g we can assume that $\sz{\set[i\in P_j]j}=k$ for every $i$ (adding elements to some $P_j$ can only increase the right-hand side of the stated inequality by \fctref{f_KL_monotone}).
Under this assumption we can apply \lemref{l_Shear}, concluding that
\m{k\tm\h{(X_i)_{i=1}^m}\le\sum_{j=1}^r\h{(X_i)_{i\in P_j}}.}
Applying \clmref{c_E2KL}
leads to
\mal{k\tm\KL{(X_i)_{i=1}^m}{(Y_i)_{i=1}^m}
&=k\h{(Y_i)_{i=1}^m}-k\h{(X_i)_{i=1}^m}\\
&\ge k\h{(Y_i)_{i=1}^m}-\sum_{j=1}^r\h{(X_i)_{i\in P_j}}\\
&=\sum_{j=1}^r\l(\h{(Y_i)_{i\in P_j}}-\h{(X_i)_{i\in P_j}}\r),}
where the last equality follows from mutual independence of \pl[Y_i] and the assumption that every $i$ belongs to exactly $k$ sets among $P_1\dc P_r$.
The result follows.}

\sect{Read-$k$ families of functions}

We prove \theoref{t_intro_tail} in this section. We first fix notations. Let $X_1\dc X_m$ be independent random variables. For $j \in [r]$, let $P_j\sbseq[m]$ be a subset and let $Y_j=f_j(\l(X_i\r)_{i\in P_j})$. We assume that $Y_1,\ldots,Y_r$ are a read-$k$ family, that is, $\sz{\set[i\in P_j]j}\le k$ for every $i\in [m]$. Let $\PR{Y_i=1}=p_i$, and let $p$ be the average of $p_1,\ldots,p_r$. Let $\eps>0$ be fixed. We will show that
\begin{equation}\label{eq:upper_bound}
\PR{Y_1+\ldots+Y_r \ge (p+\eps) r} \le e^{-\KL{p+\eps}{p} \cdot r/k}
\end{equation}
and
\begin{equation}\label{eq:lower_bound}
\PR{Y_1+\ldots+Y_r \le (p-\eps) r} \le e^{-\KL{p-\eps}{p} \cdot r/k}.
\end{equation}

We first note that it suffices to prove the theorem in the case where each $X_i$ is uniform over a finite set $A_i$. This is since we can assume w.l.o.g that each $X_i$ is a discrete random variable, and any discrete distribution can be approximated to arbitrary accuracy by a function of a uniform distribution over a large enough finite set.
We denote by $A \deq A_1 \times \ldots \times A_m$ the set of all possible inputs, and let $\mu$ denote the uniform distribution over $A$.

We start by proving~\eqref{eq:upper_bound}. For $t=(p+\eps)r$ let us denote by $A^{\ge t}$ the set of inputs for which $\sum f_i \ge t$,
\m{A^{\ge t}\deq
\set[\sum_{j=1}^r f_j(a_1\dc a_m)\ge t]
{(a_1\dc a_m)\in A_1\dtm A_m}.}
We denote by $\mu^{\ge t}$ be uniform distribution over $A^{\ge t}$. We next define restrictions of these distributions to the sets $P_1,\ldots,P_m$.
For a set $P_j \subseteq [m]$ we denote by $\mu_{j}$ the restriction of $\mu$ to the coordinates of $P_j$ (it is the uniform distribution over $\prod_{i \in P_j} A_i$), and by $\mu_{j}^{\ge t}$ the restriction of $\mu^{\ge t}$ to the coordinates of $P_j$.

We wish to upper bound the probability that $\sum f_j\ge t$. Equivalently, we wish to upper bound $|A^{\ge t}|/|A|$.
Taking logarithms, this is equal to $\h {\mu^{\ge t}} - \h{\mu}$. By \fctref{f_KLu}, this is equivalent to $\KL{\mu^{\ge t}}{\mu}$. That is,
\begin{equation}
\PR[\mu]{\sum f_j \ge t} = \frac{|A^{\ge t}|}{|A|} = \exp\l(\h{\mu^{\ge t}}-\h{\mu}\r) = \exp\l(-\KL{\mu^{\ge t}}{\mu}\r).
\end{equation}
So, we need to obtain a lower bound on $\KL{\mu^{\ge t}}{\mu}$. Naturally, to do that we will use \crlref{c_Shear}. We thus have
\begin{equation}\label{eq:upper_bound_by_KL}
\PR[\mu]{\sum f_j \ge t} \le \exp\l(-\frac{1}{k} \sum_{j=1}^r \KL{\mu_j^{\ge t}}{\mu_j}\r).
\end{equation}
Recall that $p_j=\PR[\mu]{f_j=1}$ is the probability that $f_j=1$ under the uniform distribution. Let us denote by
$q_j=\PR[\mu^{\ge t}]{f_j=1}$ the probability that $f_j=1$ conditioned on $\sum f_j \ge t$. By \fctref{f_KL_monotone} (using $\phi=f_j$) we have that $\KL{\mu_j^{\ge t}}{\mu_j} \ge \KL{q_j}{p_j}$, hence
\begin{equation}\label{eq:upper_bound_qj}
\PR[\mu]{\sum f_j \ge t} \le \exp(-\frac{1}{k} \sum_{j=1}^r \KL{q_j}{p_j}).
\end{equation}
By convexity of Kullback-Leibler divergence (\fctref{f_KL_convex}) we have that
$$
\frac{1}{r}\sum_{j=1}^r \KL{q_j}{p_j} \ge \KL{q}{p},
$$
where $q=\frac{1}{r} \sum_{j=1}^r q_j$ and $p=\frac{1}{r} \sum_{j=1}^r p_j$. To conclude, recall that $q_j$ is the probability that $f_j=1$ given that $f_1+\ldots+f_r \ge t$. Hence the sum $\sum_{j=1}^r q_j$ is the expected number of $f_j$ for which $f_j=1$, which by definition is at least $t$. Hence $q \ge t/r = p+\eps$ and $\KL{q}{p} \ge \KL{p+\eps}{p}$ by \fctref{f_KL_increasing}.
We have thus shown that
$$
\PR[\mu]{f_1+\ldots+f_r \ge (p+\eps)r} \le \exp(-\KL{p+\eps}{p} \cdot r/k).
$$
which establishes~\eqref{eq:upper_bound}.

The proof of~\eqref{eq:lower_bound} is very similar. Define for $t=(p-\eps) r$ the set $A^{\le t}$ as
\m{A^{\le t}\deq
\set[\sum_{j=1}^r f_j(a_1\dc a_m)\le t]
{(a_1\dc a_m)\in A_1\dtm A_m}.}
Let $\mu^{\le t}$ be the uniform distribution over $A^{\le t}$, and let $\mu_{j}^{\le t}$ be its projection to $(X_i)_{i \in P_j}$. Then analogously to~\eqref{eq:upper_bound_by_KL} we have
\begin{equation}\label{eq:lower_bound_by_KL}
\PR[\mu]{\sum f_j \le t} \le \exp\l(-\frac{1}{k} \sum_{j=1}^r \KL{\mu_j^{\le t}}{\mu_j}\r).
\end{equation}
Let $q'_j=\PR[\mu^{\le t}]{f_j=1}$ be the probability that $f_j=1$ conditioned on $\sum f_j \le t$. Then analogously to~\eqref{eq:upper_bound_qj} we have
\begin{equation}\label{eq:lower_bound_qj}
\PR[\mu]{\sum f_j \le t} \le \exp(-\frac{1}{k} \sum_{j=1}^r \KL{q'_j}{p_j}).
\end{equation}
Let $q'=\frac{1}{r} \sum_{j=1}^r q'_j$. Then by convexity of Kullback-Leibler divergence, we have 
$$
\frac{1}{r}\sum_{j=1}^r \KL{q_j}{p_j} \ge \KL{q'}{p}.
$$
Now, $q' \le t/r=p-\eps$ by definition, hence by \fctref{f_KL_decreasing} $\KL{q'}{p} \ge \KL{p-\eps}{p}$. So, we conclude that
$$
\PR[\mu]{f_1+\ldots+f_r \le (p-\eps)r} \le \exp(-\KL{p-\eps}{p} \cdot r/k).
$$
which establishes~\eqref{eq:lower_bound}.

\subsection*{Acknowledgments}

We are grateful to Noga Alon, Russell Impagliazzo and Pavel Pudl\'ak for helpful discussions.
DG acknowledges support by ARO/NSA under grant W911NF-09-1-0569. SL acknowledges support by NSF grant DMS-0835373.
MS acknowledges support by NSF Grant DMS-0832787.

\bibliography{atrk}

\MyComment{Look for ...-s}

\MyComment{Spell-check}

\end{document}